\documentclass[journal]{IEEEtran}
\usepackage{amsmath,amsfonts}
\usepackage{algorithmic}
\usepackage{algorithm}
\usepackage{array}
\usepackage[caption=false,font=normalsize,labelfont=sf,textfont=sf]{subfig}
\usepackage{textcomp}
\usepackage{stfloats}
\usepackage{url}
\usepackage{verbatim}
\usepackage{graphicx}
\usepackage{cite}

\usepackage{bbm}
\usepackage{multirow}

\newcommand{\bfA}{{\bf A}}
\newcommand{\bfE}{{\bf E}}
\newcommand{\bfQ}{{\bf Q}}
\newcommand{\bfT}{{\bf T}}
\newcommand{\bfe}{{\bf e}}
\newcommand{\bfJ}{{\bf J}}
\newcommand{\bfs}{{{\bf s}}}
\newcommand{\bfx}{{\bf x}}
\newcommand{\bfy}{{\bf y}}
\newcommand{\bfz}{{\bf z}}
\newcommand{\bfr}{\hat{\bf r}}
\newcommand{\bfw}{{\bf w}}
\newcommand{\be}{\begin{equation}}
\newcommand{\ee}{\end{equation}}
\newcommand{\beq}{\begin{eqnarray}}
\newcommand{\eeq}{\end{eqnarray}}

\usepackage{color}

\begin{document}

\title{\huge Weight Selection for Pattern Control of Paraboloidal  Reflector Antennas with Reconfigurable Rim Scattering}

\author{R. Michael Buehrer, {\it IEEE Fellow} and S.W. Ellingson {\it Senior Member, IEEE}\thanks{The authors are with {\em Wireless @ Virginia Tech} the Bradley Dept. of Electrical and Computer Engineering, Virginia Tech, Blacksburg, VA 24060, USA.}\thanks{Corresponding author: buehrer@vt.edu}}

\maketitle

\begin{abstract}
It has been recently demonstrated that modifying the rim scattering of  a paraboloidal reflector antenna through the use of reconfigurable elements along the rim facilitates sidelobe modification including cancelling sidelobes.  In this work we investigate techniques for determining the unit-magnitude weights (i.e., weights which modify the phase of the scattered signals) to accomplish sidelobe cancellation at arbitrary angles from the reflector axis.  Specifically, it is shown that despite the large search space and the non-convexity of the cost function, weights can be found with reasonable complexity which provide significant cancellation capability.  First, the optimal weights without any magnitude constraints are found.  Afterwards, algorithms are developed for determining the unit-modulus weights with both quantized and unquantized phases.  Further, it is shown that weights can be obtained that both cancel sidelobes while providing a constant main lobe gain.  A primary finding is that sufficiently deep nulls are possible with essentially no change in the main lobe with practical (binary or quaternary) phase-only weights.
\end{abstract}


\section{Introduction}
\label{sec:intro}
Radio astronomy is an example application that relies on large reflector antennas with high gain to receive weak signals \cite{rohlfs2013tools,CondonRansom+2016,ellingson2015antennas}.  However,  with the high gains obtainable with large reflectors, these applications are still vulnerable to interference via sidelobes. This problem can be ameliorated by sidelobe modiﬁcation or better yet,  sidelobe cancellation.  The latter involves placing a pattern null so as to reject the interference, ideally without impacting the main lobe gain. The traditional approach to sidelobe canceling is to use an array  of feeds \cite{bird2015fundamentals}.  The short-coming of such an approach is that it introduces greater aperture blockage and dynamic variability in the gain of the main lobe.

Recently,  another approach for cancelling (or generally modifying) the sidelobe(s) of a reflector antenna pattern has been proposed  that  uses reconﬁgurable rim scattering \cite{Ellingson21}.  In that work it was shown that by using reconfigurable segments along the rim of a refletor antenna that introduce phase shifts to the reflected signal, sidelobe patterns can be altered and even cancelled.  That work examined  prime focus-fed circular axisymmetric paraboloidal reﬂectors,  although the concept is not limited to that type of system.  Specifically, it was shown in \cite{Ellingson21} that as long as  sufficient surface area along the rim is reconfigurable, sidelobe cancellation is possible.  Additionally, the work showed that cancellation did not  require continuously variable phase, but was in fact possible using quantized phase shifts  (viz., binary or quaternary).  A rudimentary example  algorithm for determining the binary weights was also described.  We shall refer to this algorithm as {\it serial search}.

In the current work we build on the idea proposed in \cite{Ellingson21} by describing algorithms for determining the weights needed to cancel sidelobes at specific angles from the reflector axis in an open-loop fashion.  More specifically, we first derive the optimal weights needed to cancel sidelobes at an arbitrary angle $\psi$.\footnote{We will define the specific meaning of $\psi$ shortly.}  Due to the large number of segments available on the reflector rim, there is a large number of degrees of freedom.  Thus, we show multiple ways that these degrees of freedom can be used to cancel sidelobes at multiple angles.  The optimal weights for this case are also derived.  However, the optimal weights prove to have two short-comings:  (a) the weights are, in general, not unit-modulus; and (b) the optimal weights cause the main lobe gain to vary as a function of the null direction.  

To overcome the latter problem, we show that by exploiting the large number of degrees of freedom, an additional constraint for the main lobe gain can be added which removes the main lobe gain variability.  The former problem (i.e., requiring segment weights with non-unit gain) is more significant since such weights would require segments which must be able to attenuate or amplify the signal while scattering.  This is clearly undesirable from a cost/complexity perspective.  To overcome this problem, we formulate a least-squares problem and utilize a version of the gradient projection algorithm to force the weights to have unit modulus.  It is shown that this algorithm can be successfully used to find weights that form one or more nulls while maintaining a constant main lobe gain. 

While the algorithm described above does achieve what we desire, it requires continuously-variable phase at each reconfigurable segment.  To overcome this requirement, we must quantize the weights at each segment to a small number of possible phase shifts.  Unfortunately, this now becomes an optimization problem with a discrete multi-dimensional search space.  The non-convex discrete set of possible weights does not lend itself to the aforementioned search algorithm.  Thus, we employ a different approach.  Specifically we apply simulated annealing and show that it does result in weights that can form nulls at arbitrary angles.  A primary finding is that sufficiently deep nulls are possible with essentially no change in the main lobe with practical (binary or quartenary) phase-only weights.

This contribution is organized as follows:  Section \ref{sec:system_model} describes the system model used in this work.  The algorithms for determining the weights for each reconfigurable segment are presented in \ref{sec:weights}.  Section \ref{sec:results} presents numerical results and conclusions are presented in Section \ref{sec:concl}.

\section{System Model}
\label{sec:system_model}
\begin{figure}
    \centering
    \vspace{1cm}
    \includegraphics[ width=0.85\columnwidth]{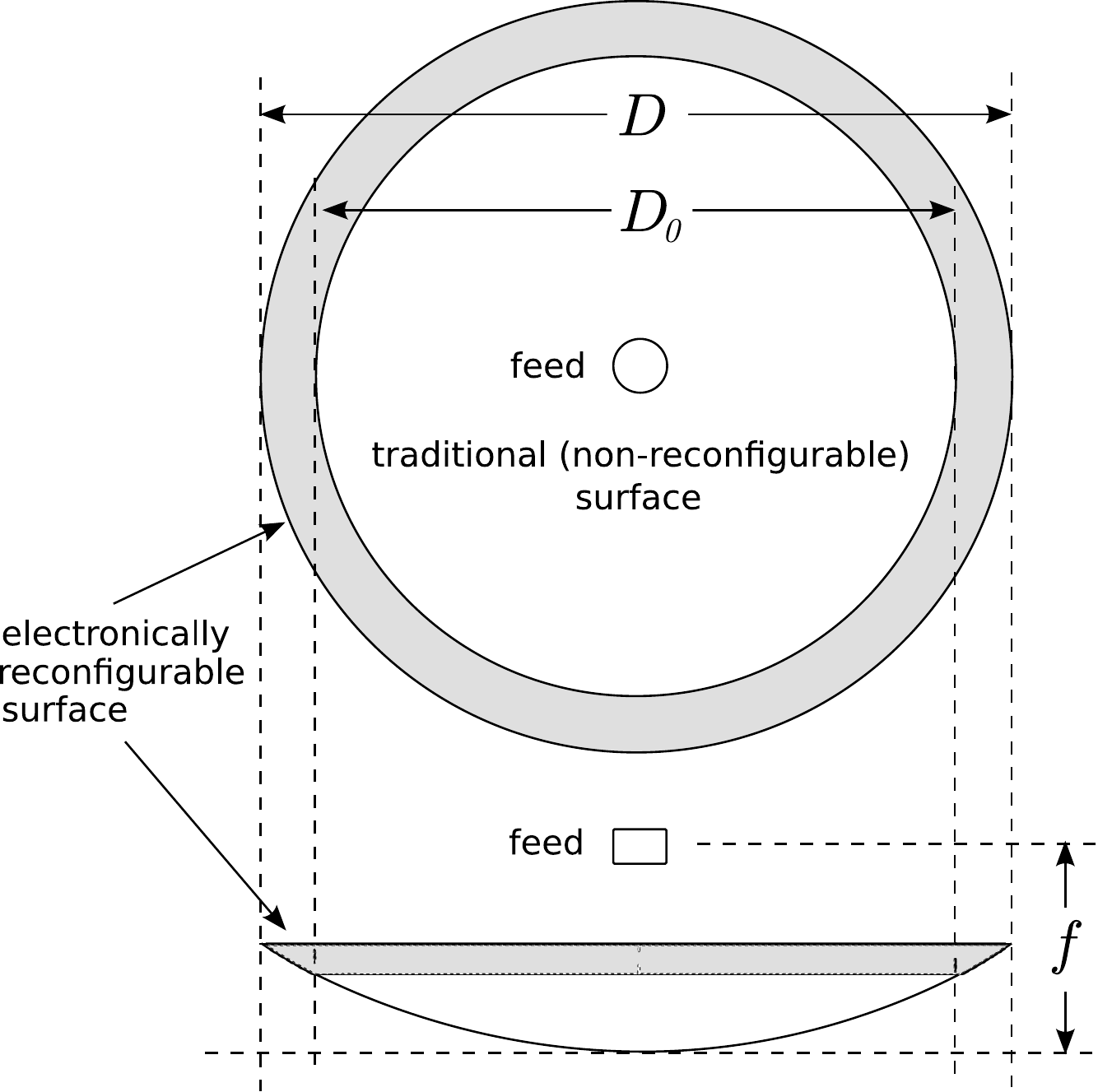}
    \includegraphics[ width=0.85\columnwidth]{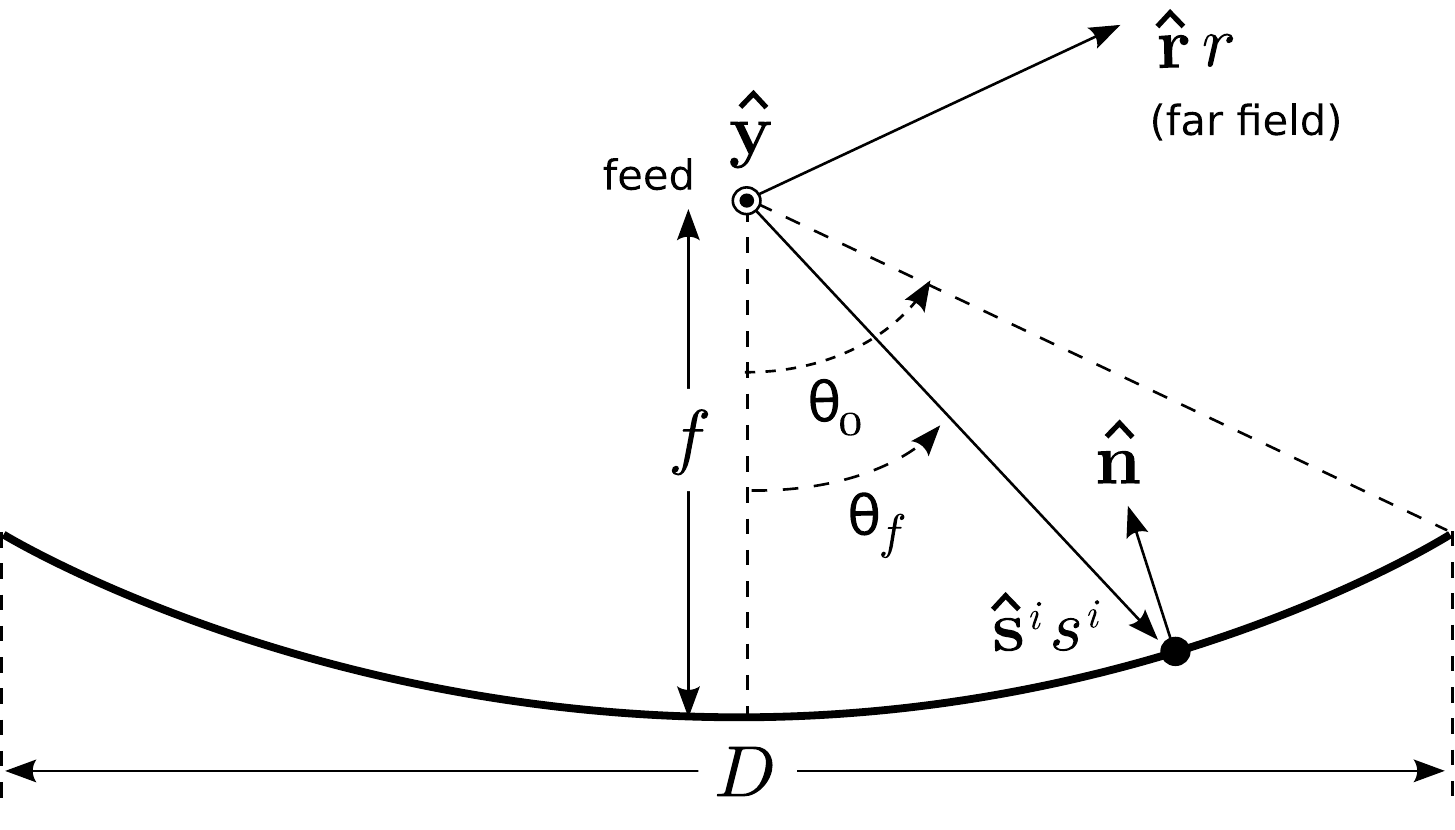}
    \caption{On-axis ( top ) and side ( middle ) views of an electronically-
reconﬁgurable rim scattering system along with the geometry for analysis (bottom) assumed in this paper.}
    \label{fig:fig0}
    \vspace{1cm}
\end{figure}

The antenna system assumed in this paper is presented in Figure \ref{fig:fig0}.  Following the development in \cite{Ellingson21} we assume the equivalence of transmit and receive patterns and calculate the transmit patterns using physical optics (PO).  The total electric field intensity $\bfE^s$ scattered by the reflector in the far-field direction $\psi$ is given by\footnote{For simplicity/clarity and with no loss of generality we will examine the antenna pattern in the H-plane.  Further, we examine the pattern versus an angle we define as $\psi$ which is the angle from the main lobe in the H-plane (i.e., $\psi=\theta$ with $\phi=90^\circ$).  Clearly the antenna pattern in general depends on both angular coordinates, but for ease of exposition we choose to stay in the H-plane in this discussion.} 
\be
\bfE^s(\psi) = \bfE^s_f(\psi) + \bfE_r^s(\psi)
\ee
where $\bfE_f^s$ is the electric field intensity due to the fixed portion of the dish
\be
\bfE_f^s(\psi) = -j\omega \mu_o \frac{e^{-j\beta r}}{4\pi r} \int_{\theta_f=0}^{\theta_1} \int_{\phi=0}^{2\pi} \bfJ_0(\bfs^i) e^{j\beta\bfr(\psi)\cdot\bfs^i}ds
\ee
where $\bfJ_0(\bfs^i)$ is the PO equivalent surface current distribution, $j=\sqrt{-1}$, $\omega=2\pi f$ is the operating frequency in rad/s, $\mu_o$ is the permeability of free space, $\beta$ is the wavenumber, $\bfr(\psi)$ points from the origin towards the field point in the direction $\psi$, $\theta_f$ is the angle measured from the reﬂector axis of rotation toward the rim with $\theta_f = \theta_1$ at the rim of the fixed portion of the dish and $\theta_f=\theta_0$ at the rim of the entire dish (see Figure \ref{fig:fig0}), $\phi$ is the angular coordinate orthogonal to both $\theta_f$ and the reﬂector axis, and $ds$ is the differential element of surface area.  See Figure \ref{fig:fig0} for more details.

The electric field intensity due to the reconfigurable portion of the dish is similarly written as
\be
\bfE_r^s(\psi) = -j\omega \mu_o \frac{e^{-j\beta r}}{4\pi r} \int_{\theta_f=\theta_1}^{\theta_o} \int_{\phi=0}^{2\pi} \bfJ_1(\bfs^i) e^{j\beta \bfr(\psi)\cdot\bfs^i}ds
\ee
where the major differences between the contributions due to the fixed and reconfigurable portions of the dish are (a) the angles of integration for the rim and (b) the PO equivalent surface current distribution.  Due to the discrete nature of the reconfigurable surface, we can write $\bfE_r^s(\psi)$ as 
\be
\bfE_r^s(\psi) = -j\omega \mu_o \frac{e^{-j\beta r}}{4\pi r} \sum_n \bfJ_1(\bfs_n^i) e^{j\beta\bfr(\psi)\cdot\bfs^i_n}\Delta s
\ee
where $\bfJ_1(\bfs_n^i)=w_n \bfJ_0(\bfs_n^i) $ is the current distribution due to the $n$th segment with complex-valued weight $w_n$.  These weights will be designed to cancel sidelobes in the H-plane co-pol pattern.  Thus, we are primarily concerned with the $y$-component of the vector $\bfE_r^s(\psi)$.  Thus, we define the complex scalar $E_r^{s,co}(\psi)$ to be the $y$-component of the vector $\bfE_r^s(\psi)$.  Also, for convenience, we can write $E_r^{s,co}(\psi)$  in terms of two $N \times 1$ dimensional arrays $\bfe_\psi$ and $\bfw$ representing the co-pol portion of the electric field intensity without the influence of the segments and the complex-valued segment gains respectively:
\be
E_r^{s,co}(\psi) = \bfe_\psi^T \bfw
\ee
where $\bfx^T$ is the transpose of the array $\bfx$, $w_n$ is the complex-valued weight applied to the $n$th reconfigurable segment and the $n$th element of $\bfe_\psi$ is 
\be
 e_{\psi,n} = \left (-j\omega \mu_o \frac{e^{-j\beta r}}{4\pi r} \bfJ_o(\bfs_n^i) e^{j\beta \bfr(\psi)\cdot\bfs^i_n}\Delta s \right ) \cdot \hat{\bf y},
\ee
 where the dot product $\left (\bfe \right )\cdot \hat{\bf y}$ selects the $y$-component of the vector $\bfe$.  Note that both one-dimensional arrays are of length $N$ where $N$ is equal to the number of reconfigurable segments placed along the rim of the dish. 

Now to cancel the sidelobe gain at angle $\psi$, we wish 
\be
E_f^{s,co}(\psi) + E_r^{s,co}(\psi) = 0
\ee
or in other words, $E_r^{s,co}(\psi) = -E_f^{s,co}(\psi)$.  Thus, we wish to find the set of weights  $\bfw$ such that
\be
\bfe_\psi^T\bfw = -E_f^{s,co}(\psi)
\label{eq:req}
\ee
In the following section we describe techniques to determine $\bfw$.

\section{Weight Selection} \label{sec:weights}
The approach to find the appropriate weights to satisfy (\ref{eq:req}) depends on the restrictions placed on $\bfw$.  First, we describe how to determine the weights with no restrictions placed on $\bfw$.  Second, we will describe techniques for finding $\bfw$ if the weights are restricted to unit-modulus values.  Finally, we describe approaches to finding weights if we further restrict the weights to be discrete.
\subsection{Unquantized Weights}
While (\ref{eq:req}) provides the requirement for creating a null at angle $\psi$, it does not provide the specific weights $\bfw$ to achieve this.
If the weights are unconstrained, to form a null at angle $\psi$ we can simply let $\bfw$ be equal to
\be
\bfw_{opt} = -E_f^{s,co}(\psi) \frac{\bfe_\psi^*}{||\bfe_\psi||^2_2}
\label{eq:opt}
\ee
where $\bfx^*$ represents a vector where each element is the complex conjugate of the corresponding element in $\bfx$ and $||\bfx||_2$ represents the 2-norm of the vector $\bfx$.  We will term these {\it optimal} weights since these weights guarantee that the response of the dish in the direction $\psi$ is zero.  Unfortunately, $\bfw_{opt}$ in general has elements with $|w_i|\neq 1$ which requires that the segments have controllable gains (i.e., can provide attenuation or gain to the scattered field). Such a requirement is undesirable from a cost and complexity perspective.  Thus, we seek to restrict the weights such that $|w_i|=1$.  This can be written as the following minimization problem
\begin{eqnarray}
     \bfw_{gp} = \min_{\bfw \in {\cal C}^N} & \left \| E_f^{s,co}(\psi) + \bfe_\psi^T\bfw  \right \|_2^2 & \label{eq:opt_single} \\
    &  s.t. \hspace{0.25cm} |w_i| = 1 & i =1, 2, \ldots N \nonumber 
\end{eqnarray}
This is a non-convex, complex-valued, constant-modulus, least squares optimization problem which is a special case of non-convex quadratically-constrained quadratic programming \cite{Luo10}.  One solution to this problem is to use semi-definite programming \cite{Luo10}.  Unfortunately, this expands the problem such that it is of dimension $N^2$.  Given that for moderate to large antennas ($D=15$ m to $D=100$ m) and GHz frequencies, $N$  can range from $10^3$ to $10^5$, such an expansion is computationally complex.  A more efficient solution is the use of {\it Gradient Projection} \cite{Tranter17}.  Gradient Projection (GP) uses standard gradient descent followed by a projection onto a convex set.  While our set of unit-modulus vectors is not a convex set, this approach has been shown to converge for projection onto the set of unit-modulus vectors \cite{Tranter17}.  Adapting GP to our problem, we have the following algorithm.  At the $k$th iteration we apply standard gradient descent to create a new vector 
\be
\tilde{\bfw}^{(k+1)} = \bfw^{(k)} + \alpha \bfe^* \left ( E_f^{s,co}(\psi) + \bfe^T\bfw^{(k)} \right )
\ee
where $\alpha = \frac{\gamma}{\lambda_{max}\left ( \bfe^* \bfe^T \right) }$ is the adaptation constant and $\gamma \in (0,1)$.  Of course, this is not guaranteed to provide unit-modulus weights.  Thus, we follow this with a projection step which projects $\tilde{\bfw}$ onto an array of weights with unit-modulus $\bfw$:
\be
\bfw^{(k+1)} = {\cal P} \left (\tilde{\bfw}^{(k+1)} \right )
\ee
where ${\cal P}(\bfx) = e^{j\angle \bfx}$ and the vector operator $\angle \bfx$ results in a second vector $\bfz$ where the $i$th element is the angle of $x_i$,i.e.,  $z_i=\angle x_i$ .  The detailed algorithm follows.

\begin{table}[h]
    \centering
    \begin{tabular}{l|l}
    \hline 
    \multicolumn{2}{c}{GP Algorithm (single constraint/null)} \\
    \hline
     1. & Initialization: $k=0$, $\alpha = \frac{\gamma}{\lambda_{max}\left ( \bfe^* \bfe^T \right) }$, $\gamma \in (0,1)$ \\
      & \hspace{2cm} $\bfw^{(0)} =e^{j\angle \left ((\bfe^*\bfe^T)^{-1}\bfe^*(-E_f^{s,co}(\psi)) \right )} $ \\
      & \\
         & Repeat \\
      2. & \hspace{1cm} ${\bf \eta}^{(k+1)} = \bfw^{(k)} - \alpha \bfe^* \left ( E_f^{s,co}(\psi) + \bfe^T\bfw^{(k)} \right )$ \\
      3. & \hspace{1cm} $\bfw^{(k+1)} = e^{j\angle {\bf \eta}^{(k+1)}}$ \\
      4. & \hspace{1cm} $k=k+1$ \\
         & Until convergence  \\
      \hline
    \end{tabular}
    %\caption{Caption}
    \label{tab:my_label}
\end{table}

\subsection{Main Lobe Variation and Nulling Multiple Angles}
In the above formulation, we chose weights to cancel the sidelobe at a single angle.  However, there may be multiple angles that we wish to null. Further, as we will show later, nulling a particular angle may cause small changes to the main lobe gain (this was also shown in \cite{Ellingson21}).  This variation  may be tolerable in some applications, but in radio astronomy it can be a significant problem\cite{CondonRansom+2016}.  Thus, we wish to constrain this variation as much as possible.  To accomplish both goals (adding nulls and restricting main lobe variation) we simply need to add these requirements to the cost function.  

For example, consider the case of $K$ desired nulls.  In this case we wish
\beq
E_f^{s,co}(\psi_1) + E_r^{s,co}(\psi_1) & = & 0 \nonumber \\ 
E_f^{s,co}(\psi_2) + E_r^{s,co}(\psi_2) & = & 0 \nonumber \\ 
& \vdots & \nonumber \\
E_f^{s,co}(\psi_K) + E_r^{s,co}(\psi_K) & = & 0 
\eeq
Or in other words we desire
\beq
 \bfe_{\psi_1}^T\bfw & = & -E_f^{s,co}(\psi_1) \nonumber \\ 
& \vdots & \nonumber \\
 \bfe_{\psi_K}^T\bfw & = & -E_f^{s,co}(\psi_K)   
\eeq
Defining 
\beq
\bfQ & = & \left [\bfe_{\psi_1}, \bfe_{\psi_2}, \ldots \bfe_{\psi_K}   \right ] \\ 
\bfy & = & -[E_f^{s,co}(\psi_1), E_f^{s,co}(\psi_2), \ldots E_f^{s,co}(\psi_K)  ]^T, 
\eeq
we have the requirement 
\be
\bfQ^T \bfw = \bfy
\ee
Using the least squares solution we obtain the optimal weights
\be
\bfw_{opt} = \bfQ \left ( \bfQ^T \bfQ\right )^{-1} \bfy
\ee

To avoid variation in the main lobe, we can include a constraint for the main lobe.  Specifically, for $\psi=0$ we require
\be
\bfe_{0}^T\bfw = \kappa
\ee
where $\kappa$ is the target constraint.  Ideally, we could set $\kappa = \bfe_{0}^T{\bf 1}_N$ where ${\bf 1}_N$ is an $N\times 1$ array of all ones which would provide the same main lobe gain as the fixed reflector.  With unconstrained weights, this is possible.  However, with weights restricted to unit modulus, satisfying both the main lobe constraint and side lobe constraints many not be possible.  One option is to ease the constraints by choosing $\kappa=0$, although this removes any contribution of the reconfigurable portion of the reflector to the main lobe.  Alternatively, we could choose $\kappa = \delta E_f^{s,co}(0)$ for some small value $\delta$ to provide some additional gain in the main lobe.  Experimentally we have found this latter approach to be successful. Thus, we can modify the vector of requirements to be
\be
\bfy = \left [\kappa,-E_f^{s,co}(\psi_1), -E_f^{s,co}(\psi_2), \ldots -E_f^{s,co}(\psi_K) \right ]^T
\ee
where $\psi_1, \psi_2, \ldots \psi_K$ are the $K$ angles at which we desired to place nulls and the first element of $\bfy$  corresponds to the constraint placed on the change to the main lobe gain.  Further, we define the matrix
\be
\bfA  = \left [ \bfe_{\psi_0},\bfe_{\psi_1}, \bfe_{\psi_2},\ldots \bfe_{\psi_K},\right ]^T
\ee
The least squares problem then becomes:
\begin{eqnarray}
   \bfw_{gp} =  \min_{\bfw \in {\cal C}^N} & \left \| \bfy + \bfA \bfw  \right \|_2^2 & \label{eq:opt_multiple} \\
    &  s.t. \hspace{0.25cm} |w_i| = 1 & i =1, 2, \ldots N \nonumber 
\end{eqnarray}

The resulting algorithm is then

\begin{table}[h]
    \centering
    \begin{tabular}{l|l}
    \hline 
    \multicolumn{2}{c}{GP Algorithm (multiple constraints)} \\
    \hline
     1. & Initialization: $k=0$, $\alpha = \frac{\gamma}{\lambda_{max}\left ( \bfA^H \bfA \right) }$, $\gamma \in (0,1)$ \\
      & $\bfw^{(0)} =e^{j\angle \left ((\bfA^H\bfA)^{-1}\bfA^H\bfy \right )} $ \\
      & \\
       & Repeat \\
      3. & \hspace{1cm} ${\bf \eta}^{(k+1)} = \bfw^{(k)} - \alpha \bfA^H \left ( \bfy + \bfA \bfw^{(k)} \right )$ \\
      4. & \hspace{1cm}  $\bfw^{(k+1)} = e^{j\angle {\bf \eta}^{(k+1)}}$ \\
      5. & \hspace{1cm}  $k=k+1$ \\
        & Until convergence  \\
      \hline
    \end{tabular}
    %\caption{Caption}
    \label{tab:my_label2}
\end{table}
\subsection{Quantized Weights}
The weights described in the previous sections have one primary disadvantage: they presume continuously-variable phase values.  In a practical implementation, it is much more reasonable that only a finite number of phase values would be available on each reconfigurable segment.  Thus, we wish to solve the single-null problem in (\ref{eq:opt_single}) or the multiple-null problem (\ref{eq:opt_multiple}) with the additional constraint that 
\be
w_i \in {\cal W}=\left \{ e^{j2\pi/M},e^{j4\pi/M}, \ldots e^{j2\pi(M-1)/M} \right \}.
\ee
where $M$ is the number of possible phase values.  The most straightforward approach is to simply search over all  vectors $\bfw\in {\cal W}$ to find the one that minimizes the cost function in (\ref{eq:opt_single}) for a single constraint or (\ref{eq:opt_multiple}) for multiple constraints.  Unfortunately, this requires a search over $M^N$ possible vectors where (as discussed above) $N$ can be on the order of $10^3$ to $10^5$ for moderate to large antennas and GHz frequencies.   For example, in the particular case we will examine $N=2756$ (see Section \ref{sec:results} and \cite{Ellingson21} for details). Thus, even using binary weights a brute-force approach results in a search over $2^{2756}$ possible values of $\bfw$ which is obviously infeasible.  

A well-known and much more efficient approach is  simulated annealing \cite{russell2002artificial}.  In this approach we initialize a parameter known as the {\it temperature} and perform a random search that becomes more greedy in nature as the temperature ``cools''.   In detail, the algorithm is as follows where the cost function $ C \left ( \bfw \right )=\left \| E_f^{s,co}(\psi) + \bfe_\psi^T\bfw  \right \|_2^2$ for a single constraint and  $ C \left ( \bfw \right )=\left \| \bfy + \bfA\bfw  \right \|_2^2$ for multiple constraints:

\begin{table}[h]
    \centering
    \begin{tabular}{l|l}
    \hline 
    \multicolumn{2}{c}{Simulated Annealing} \\
    \hline
     1. & Initialization: $k=0$, $\bfT=\left [ 1, \frac{1}{2},\frac{1}{3},\ldots, \frac{1}{T} \right ]$ \\
      & \hspace{1.5cm} $\bfw^{(0)}$ randomly chosen from ${\cal W}^N$ \\
         & Repeat \\
      3. & \hspace{1cm} $\bfw^{(k+1)} = \bfw^{(k)}$\\
      4. & \hspace{1cm} $n=\lceil rand*N \rceil$,$m=\lceil rand*M \rceil$ \\
      5. & $\hspace{1cm} \bfw^{(k+1)}_n={\cal W}(m)$ \\
      6. & $\hspace{1cm}\Delta E = C \left ( \bfw^{(k)}\right )-C \left ( \bfw^{(k+1)}\right ) $ \\
      7. & \hspace{1cm}if $\Delta E \geq 0$, $\bfw^{(k+1)}=\bfw^{(k)}$ \\
      8. & \hspace{1cm} else  $\bfw^{(k+1)}=\bfw^{(k+1)}$  with probability $p=\min \left(1, e^{\Delta E \bfT(k)}\right )$  \\
      9. & \hspace{1cm} $k=k+1$ \\
       & Until $k=T$  \\
      \hline
    \end{tabular}
    %\caption{Caption}
    \label{tab:my_label3}
\end{table}

%\pagebreak

Describing the above algorithm, we start by initializing a temperature cooling schedule.  In this work, for a series of $T$ time steps, we use the cooling schedule $\bfT=\left [ 1, \frac{1}{2},\frac{1}{3},\ldots, \frac{1}{T} \right ]$. At each time step we perform a local search where we randomly choose a neighbor state.  We define a neighbor as a vector with one element being different.  Thus, we randomly choose the $n$th weight element and randomly change the value to one of the other $M$ possible weights. We then compare the cost (denoted ``energy'' in simulated annealing) of each weight vector.  If the new weight vector has smaller cost (i.e., closer to the specified constraints), the weight vector is changed to this new weight vector.  However, if the new weight vector has a larger cost, with probability $p$ we keep the new weight vector despite its higher cost, and with probability $1-p$ we maintain the old weight vector.  In this way, local minima are avoided.  The key is that the probability is tied to the cooling temperature which decreases with each time step.

\section{Results}\label{sec:results}
\begin{figure}
    \centering
    \vspace{1cm}
    \includegraphics[trim={5cm 8cm 5cm 10cm}, width=0.85\columnwidth]{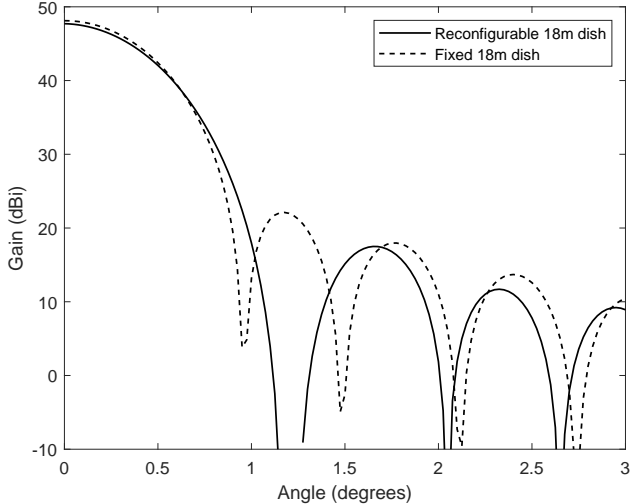}
    \caption{H-plane Co-Pol Pattern for Standard (fixed) 18m Dish and Reconfigurable 18m Dish with 0.5m Reconfigurable Rim ($\psi=1.25^o$)}
    \label{fig:fig1}
    \vspace{1cm}
\end{figure}

To demonstrate the performance of the above algorithms, in this section we provide numerical results.  We will assume a $D = 18$ m paraboloidal reflector operating at 1.5 GHz where the outer
0.5 m of the reﬂector surface consists of 2756 contiguous reconfigurable segments.  Each segment is a square ﬂat plate conformal to the  paraboloidal surface having side length $0.5\lambda$ ({\it i.e.}, an area $\Delta s = 0. 25\lambda^2$).  The feed is modeled as an electrically-short electric dipole with field additionally modiﬁed by the factor $(\cos \theta_f )^q$ ,where $q$ controls the directivity of the feed.  Setting $q = 1$ yields edge illumination (i.e., ratio of ﬁeld intensity in the direction of the rim to the ﬁeld intensity in the direction of the vertex), to approximately -11 dB, yielding aperture efﬁciency of about $81.5$\%.  Figure \ref{fig:fig1} presents the H-plane co-pol pattern of an 18 m fixed reflector (i.e., without the reconfigurable segments or equivalently with the reconfigurable segments set to $w_i=1, \forall i$).  Also plotted is the same pattern for the reconfigurable reflector with the optimal weights (i.e., infinite quantization, and no unit modulus constraint) defined in (\ref{eq:opt}).  The weights are set to place a null at $\psi = 1.25^\circ$ which is directly on the peak of the first sidelobe in the fixed reflector pattern.  We can see from the figure that the sidelobe at $\psi=1.25^\circ$ is indeed cancelled.  However, the main lobe gain is reduced from 48.1 dBi (for the fixed 18m dish) to 47.7 dBi for the reconfigurable dish.   

\begin{figure}
    \centering
    \includegraphics[trim={5cm 8cm 5cm 10cm}, width=0.85\columnwidth]{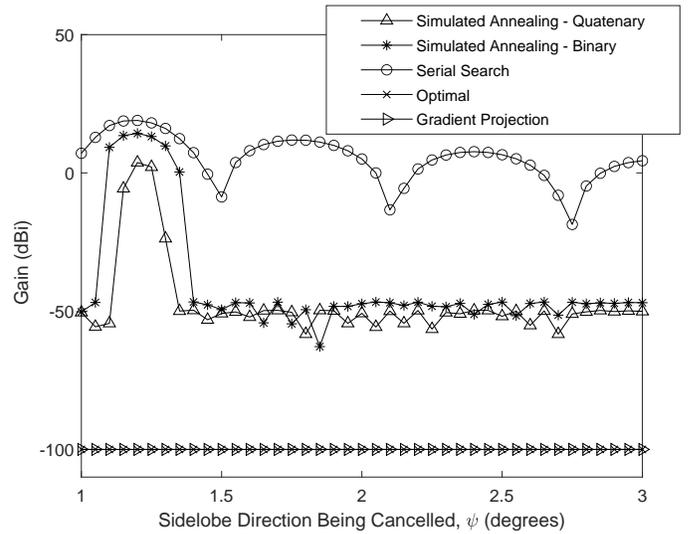}
    \caption{H-plane Co-Pol Pattern Gain $G(\psi)$ when Placing a Null at Angle $\psi$ using the Proposed Algorithms and the Serial Search Approach for Binary Weights in \cite{Ellingson21}.  Note that the gains for ``optimal'' and ``Gradient Projection'' were measured as zero to the within the limits of machine precision.  Thus, to visualize these values, we simply mapped them to -100 dBi for plotting purposes. }
    \label{fig:optimal_null}
\end{figure}

\begin{figure}
    \centering
    \vspace{1cm}
    \includegraphics[trim={5cm 8cm 5cm 10cm}, width=0.85\columnwidth]{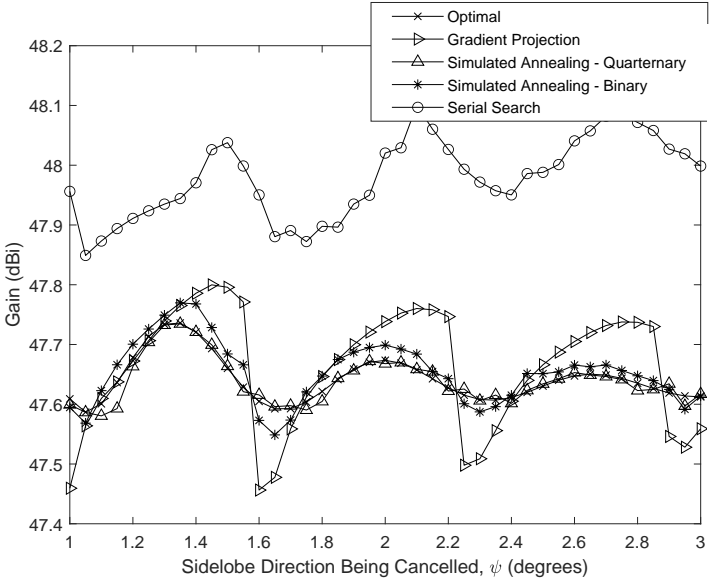}
    \caption{Main Lobe Gain for H-plane Co-Pol Pattern ($G(0)$) when Cancelling Sidelobe at $\psi$ using the Proposed Algorithms and the Serial Search Approach for Binary Weights in \cite{Ellingson21}}
    \label{fig:main_lobe}
\end{figure}

The pattern in Figure \ref{fig:fig1} resulted from weights generated to place a null at $\psi=1.25^\circ$.  As a result, the gain at $\psi=1.25^\circ$ is essentially zero.\footnote{In calculating the gains at sidelobes which were nulled, with optimal weights the gains were measured as zero to the within the limits of machine precision.  Thus, to visualize these values, we simply mapped them to -100 dBi for plotting purposes.}  This level of cancellation can be accomplished at any angle outside the main beam as shown in Figure \ref{fig:optimal_null}. Specifically, in Figure \ref{fig:optimal_null} we plot the H-plane co-pol gain achieved at $\psi$ (we will refer to this as $G(\psi)$) when the weights are generated using (\ref{eq:opt}) to place a null at $1^\circ\leq \psi \leq 3^\circ$ (see plot labeled ``optimal'').  Note that this is not a pattern, but rather the resulting null depth when attempting to place a null at $\psi$.    It can be seen that the optimal weights provide a gain of  zero to within the precision of the machine for any angle between $1^\circ$ and $3^\circ$,  Similarly, the resulting main lobe gain when cancelling a single sidelobe at angle $\psi$ is shown in Figure \ref{fig:main_lobe}.  It can be seen that the main lobe gain can vary by over 0.1 dB when cancelling with the optimal weights.  As discussed above, this variation can be problematic.  

Further, and as also discussed above, the optimal weights as determined by (\ref{eq:opt}) are not unit modulus and thus would require reconfigurable segments with controllable gain.  The gain achieved at the null direction when using the weights defined by the least-squares problem described in (\ref{eq:opt_single}) and determined using the Gradient Projection algorithm are also plotted in Figure   \ref{fig:optimal_null} (weights labeled ``Gradient Projection'').  It can be seen that the unit-modulus weights also achieve a gain of essentially zero. The corresponding mainlobe gain is also plotted in Figure  \ref{fig:main_lobe}.  The additional constraint of the unit modulus causes more mainlobe variation as the angle of sidelobe cancellation ($\psi$) is changed.  Specifically, the mainlobe gain will vary by approximately 0.3 dB  $\approx 7\%$ which may be significant in radio astronomy applications.  We will address this limitation shortly.  Of course, these weights, while unit modulus, are not quantized and thus require the segments to have continuously-variable phase.  

The null depth achievable when using binary ($M=2$) weights (via the serial search algorithm described in \cite{Ellingson21} and the simulated annealing algorithm) or quaternary weights (via the simulated annealing algorithm) are also shown in Figure \ref{fig:optimal_null}.  When using the serial search approach from \cite{Ellingson21}, the binary weights clearly get stuck in a local minimum that limits the null achievable.  The minimum gains are in the range of 0-5 dBi which is a reduction in fixed pattern gain of only  5-15 dB. On the other hand, when using simulate annealing, the quantized weights can push the sidelobe gains down to approximately -50 dBi, with the exception of a $0.2^\circ$ range around the maximum sidelobe peak at $\psi=1.25^\circ$.  In that range, there is insufficient granularity in the weights to completely null the power from the fixed portion of the dish.  On the other hand, the main lobe gain has roughly 0.2 dB of variation when using simulated annealing and 0.3 dB of variation when using the serial search, although the latter provides a higher main lobe gain by nearly 0.5 dB.  Note that this difference in main lobe gain can be reduced by changing the main lobe constraint.

\begin{figure}
    \centering
    \vspace{1cm}
    \includegraphics[trim={5cm 8cm 5cm 10cm}, width=0.85\columnwidth]{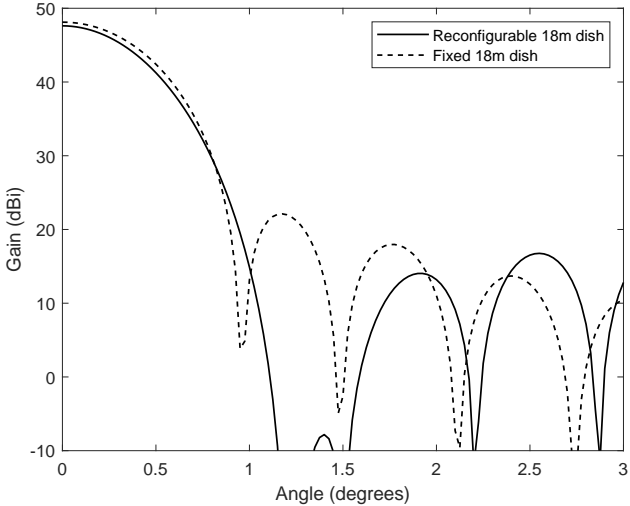}
    \caption{H-plane Co-Pol Pattern for Standard (fixed) 18m Dish and Reconfigurable 18m Dish with 0.5m Reconfigurable Rim (main lobe control, $\psi=1.25^\circ, 1.5^\circ$)}
    \label{fig:fig4}
\end{figure}

To demonstrate the performance of the algorithms to create multiple nulls, we first examine the continuously-variable phase case using gradient projection to solve the least squares problem of (\ref{eq:opt_multiple}).  As an example, we applied the algorithm to two angles: $\psi=1.25^o$ and $\psi=1.5^o$.  The resulting pattern is plotted in Figure \ref{fig:fig4} along with the pattern of an 18m dish with a fixed pattern.  We can see that the reconfigurable rim provides nulls at both angles, as desired.  There is a small loss in main lobe gain ($\sim$0.1 dB), but the two nulls are sufficiently deep.

\begin{figure}
    \centering
    \vspace{1cm}
    \includegraphics[trim={5cm 8cm 5cm 10cm}, width=0.85\columnwidth]{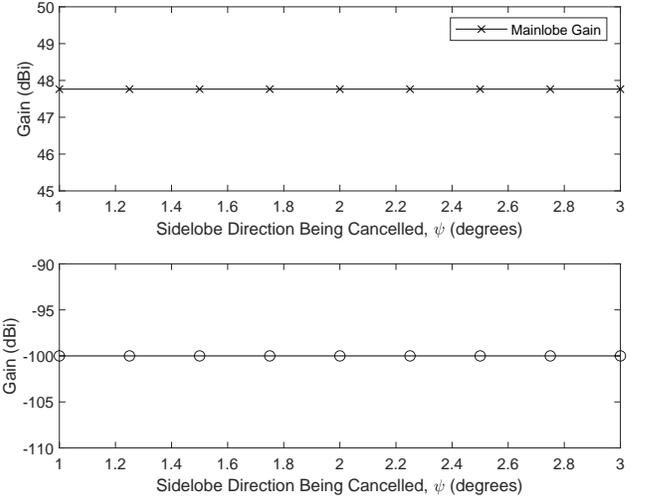}
    \caption{Main Lobe Gain of H-plane Co-Pol Pattern ($G(0)$) (top) and Sidelobe Gain ($G(\psi)$) (bottom) when Cancelling Sidelobe at $\psi$ while Enforcing A Main Lobe Constraint using Gradient Projection. Note that the sidelobe gains were measured as zero to the within the limits of machine precision.  Thus, to visualize these values, we simply mapped them to -100 dBi for plotting purposes. }
    \label{fig:fig5}
\end{figure}

As a last examination of the performance or the algorithms, we now examine the ability of the approach to eliminate main lobe gain variation (i.e., mitigate the variation seen in Figure \ref{fig:main_lobe}.  First, we use the gradient projection algorithm (as described above) with two constraints.  The first constraint is to maintain the main lobe gain at a constant value while the second constraint is to cancel the sidelobe at the angle $\psi$.  The results are shown in Figure \ref{fig:fig5} which plot the gain of the main lobe (top) and cancelled sidelobe (bottom) as the sidelobe angle $\psi$ is varied.  Again, this is not an antenna pattern but a plot of two specific gain values for different patterns as the direction of sidelobe cancellation is varied.  It can be observed that the main lobe gain is held constant at 47.8 dBi regardless of the angle of the null being created.  At the same time, to within the precision of the machine, the gain at the null direction can be set to zero.

\begin{figure}
    \centering
    \vspace{1cm}
    \includegraphics[trim={5cm 8cm 5cm 10cm}, width=0.85\columnwidth]{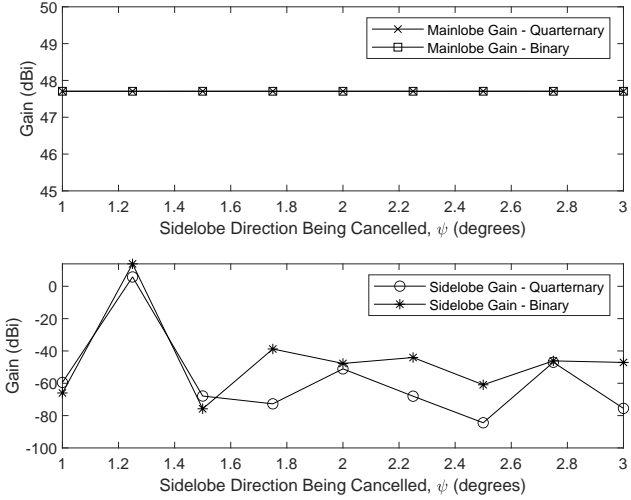}
    \caption{Main Lobe Gain $G(0)$ (top) and Sidelobe Gain (bottom) when Cancelling Sidelobe at $\psi$ while Enforcing a Main Lobe Constraint using Simulated Annealing and Quantized Weights ($M=2,4$) }
    \label{fig:fig6}
\end{figure}

\begin{figure}
    \centering
    \vspace{1cm}
    \includegraphics[trim={5cm 8cm 5cm 10cm}, width=0.85\columnwidth]{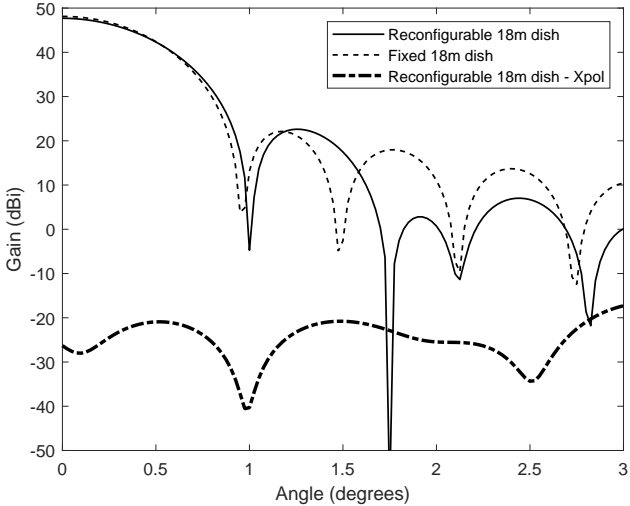}
    \caption{H-plane Co-Pol Pattern for Standard (fixed) 18m Dish and H-plane Co-Pol and Cross-Pol Patterns for Reconfigurable 18m Dish with 0.5m Reconfigurable Rim and Quaternary Weights (main lobe control, $\psi=1.75^\circ$)}
    \label{fig:fig7}
\end{figure}

If quantized weights are used, the results are not quite as good as shown in Figure \ref{fig:fig6}.  While the main lobe gain is still maintained with strong consistency, the null depth is not as low as with  continuously-variable phase weights.  This is certainly expected based on the results of Figure \ref{fig:optimal_null}.  The quantized weights do not have the same capability to create deep nulls.  This is due to the fact that quantization limits the solution space and thus weights may not exist that can accomplish both constant main lobe gain and providing a deep null at a specific direction.  This is particularly true around the largest sidelobe at $1.25^o$ as was shown above.  As an example quaternary (i.e., $M=4$) weights were generated for main lobe control and a null at $\psi=1.75^\circ$.  The resulting the H-plane co-pol and cross-pol patterns are plotted in Figure \ref{fig:fig7} along with the fixed reflector co-pol pattern.  As expected, a deep null is placed at $1.75^\circ$ while maintaining the desired main lobe gain.  Further, we can see that while the weights do degrade the cross-pol pattern  (which is zero in the H-plane in the non-reconfigurable system) the cross-pol gain does not exceed -18 dBi. Further, it is possible that with proper constraints, this cross-pol degradation could also be controlled.

\section{Conclusions}
\label{sec:concl}

In this paper we have described multiple techniques for determining optimal or near-optimal weights for creating nulls in the pattern of a prime focus-fed circular axisymmetric paraboloidal reﬂector antenna.  The general approach applies to a larger class of reflector antennas, but the discussion was limited to this specific type for demonstration purposes.  It was shown that if segments placed on the rim of the reflector are capable of imparting both amplitude and phase adjustment, creating nulls at arbitrary directions of arrival is possible.  Multiple nulls can be created and the gain of the main lobe can also be simultaneously held fixed.  

More importantly it was shown that more practical unit-modulus weights can be found using a least-squares approach based on the projection gradient algorithm.  These weights while not requiring gain control (i.e., they require phase shifts only) and capable of placing a null at any angle, unfortunately require continuously-variable phase.  However, it was also shown that deep nulls can be created using quantized unit modulus weights which require only binary or quaternary phase values using simulated annealing.  Further, it was shown that the approach can also be used to simultaneously create a null at a specific angle while maintaining a constant main lobe gain.  These approaches require knowledge of the antenna pattern and are calculated ``open-loop''.  Future work includes developing approaches for controlling the shape of the main lobe, constraining the sidelobe levels over a particular sector, and closed-loop adaptation that would not require exact knowledge of the antenna pattern.
\section*{Acknowledgement}
This material is based upon work supported in part by the National Science Foundation under Grant AST 2128506.
\bibliographystyle{IEEEtran}
\bibliography{refs}

\begin{thebibliography}{1}
\providecommand{\url}[1]{#1}
\csname url@samestyle\endcsname
\providecommand{\newblock}{\relax}
\providecommand{\bibinfo}[2]{#2}
\providecommand{\BIBentrySTDinterwordspacing}{\spaceskip=0pt\relax}
\providecommand{\BIBentryALTinterwordstretchfactor}{4}
\providecommand{\BIBentryALTinterwordspacing}{\spaceskip=\fontdimen2\font plus
\BIBentryALTinterwordstretchfactor\fontdimen3\font minus
  \fontdimen4\font\relax}
\providecommand{\BIBforeignlanguage}[2]{{%
\expandafter\ifx\csname l@#1\endcsname\relax
\typeout{** WARNING: IEEEtran.bst: No hyphenation pattern has been}%
\typeout{** loaded for the language `#1'. Using the pattern for}%
\typeout{** the default language instead.}%
\else
\language=\csname l@#1\endcsname
\fi
#2}}
\providecommand{\BIBdecl}{\relax}
\BIBdecl

\bibitem{rohlfs2013tools}
K.~Rohlfs and T.~L. Wilson, \emph{Tools of radio astronomy}.\hskip 1em plus
  0.5em minus 0.4em\relax Springer Science \& Business Media, 2013.

\bibitem{CondonRansom+2016}
\BIBentryALTinterwordspacing
J.~J. Condon and S.~M. Ransom, \emph{Essential Radio Astronomy}.\hskip 1em plus
  0.5em minus 0.4em\relax Princeton University Press, 2016. [Online].
  Available: \url{https://doi.org/10.1515/9781400881161}
\BIBentrySTDinterwordspacing

\bibitem{ellingson2015antennas}
S.~Ellingson, ``Antennas in radio telescope systems,'' \emph{Handbook of
  Antenna Technologies" ed. by Zhi Ning Chen; Singapore: Springer Singapore},
  pp. 1--21, 2015.

\bibitem{bird2015fundamentals}
T.~S. Bird, \emph{Fundamentals of aperture antennas and arrays: from theory to
  design, fabrication and testing}.\hskip 1em plus 0.5em minus 0.4em\relax John
  Wiley \& Sons, 2015.

\bibitem{Ellingson21}
S.~Ellingson and R.~Sengupta, ``Sidelobe modification for reflector antennas by
  electronically reconfigurable rim scattering,'' \emph{IEEE Antennas and
  Wireless Propagation Letters}, vol.~20, no.~6, pp. 1083--1087, 2021.

\bibitem{Luo10}
Z.-q. Luo, W.-k. Ma, A.~M.-c. So, Y.~Ye, and S.~Zhang, ``Semidefinite
  relaxation of quadratic optimization problems,'' \emph{IEEE Signal Processing
  Magazine}, vol.~27, no.~3, pp. 20--34, 2010.

\bibitem{Tranter17}
J.~Tranter, N.~D. Sidiropoulos, X.~Fu, and A.~Swami, ``Fast unit-modulus least
  squares with applications in beamforming,'' \emph{IEEE Transactions on Signal
  Processing}, vol.~65, no.~11, pp. 2875--2887, 2017.

\bibitem{russell2002artificial}
S.~Russell and P.~Norvig, \emph{Artificial Intelligence: A Modern
  Approach}.\hskip 1em plus 0.5em minus 0.4em\relax Pearson, 2002.

\end{thebibliography}

\end{document}